\documentclass{article}

\usepackage[nonatbib,final]{neurips_2022_ml4ps}
\usepackage{xcolor}
\definecolor{darkpurple}{rgb}{0.5,0,0.5}
\definecolor{mydarkgrey}{rgb}{0.27, 0.27, 0.27}
\usepackage{wrapfig}
\usepackage[backend=biber,
    style=numeric-comp ,
    maxbibnames=500, 
    natbib=true,
    doi=true,
    sorting=none,
  ]{biblatex}
\usepackage[page,header]{appendix}
\usepackage{titletoc}
 \if False
 \DeclareFieldFormat{bbx@introcite}{\textcolor{mydarkgrey}{\mkbibbold{\mkbibbrackets{#1}}}}
\DeclareDelimFormat[bbx@introcite]{nameyeardelim}{\addcomma\space}
\UndeclareInnerCiteDelims{bbx@introcite}

\DeclareFieldFormat{biblabeldate}{#1}
\DeclareDelimFormat[bib]{nameyeardelim}{\addperiod\space}

\DeclareFieldFormat{biblistlabeldate}{#1}
\DeclareDelimFormat[bib]{nameyeardelim}{\addperiod\space}

\DeclareFieldFormat{biblistlabeldate}{#1}
\DeclareDelimFormat[bib]{nameyeardelim}{\addperiod\space}

\fi
\if False
\DeclareFieldFormat{citehyperref}{%
  \DeclareFieldAlias{bibhyperref}{noformat}
  \bibhyperref{#1}}

\DeclareFieldFormat{textcitehyperref}{%
  \DeclareFieldAlias{bibhyperref}{noformat}
  \bibhyperref{%
    #1%
    \ifbool{cbx:parens}
      {\bibcloseparen\global\boolfalse{cbx:parens}}
      {}}}

\savebibmacro{cite}
\savebibmacro{textcite}

\renewbibmacro*{cite}{%
  \printtext[citehyperref]{%
    \restorebibmacro{cite}%
    \usebibmacro{cite}}}

\renewbibmacro*{textcite}{%
  \ifboolexpr{
    ( not test {\iffieldundef{prenote}} and
      test {\ifnumequal{\value{citecount}}{1}} )
    or
    ( not test {\iffieldundef{postnote}} and
      test {\ifnumequal{\value{citecount}}{\value{citetotal}}} )
  }
    {\DeclareFieldAlias{textcitehyperref}{noformat}}
    {}%
  \printtext[textcitehyperref]{%
    \restorebibmacro{textcite}%
    \usebibmacro{textcite}}}
    
\fi

\definecolor{antiquefuchsia}{rgb}{0.57, 0.36, 0.51}
\definecolor{cadetblue}{rgb}{0., 0.62, 0.63}
\definecolor{brightmaroon}{rgb}{0.76, 0.13, 0.28}
\definecolor{mymaroon}{rgb}{0.62, 0.05, 0.19}
\definecolor{mymaroon2}{cmyk}{0, 0.739, 0.614, 0.368}
\definecolor{darkpurple}{rgb}{0.5,0,0.5}
\definecolor{mydarkgrey}{rgb}{0.27, 0.27, 0.27}
\definecolor{mymagenta}{rgb}{0.8125, 0, 0.8125}
\usepackage{hyperref}

\hypersetup{colorlinks,urlcolor=darkpurple, citecolor=mymaroon, linkcolor=magenta}

\usepackage[utf8]{inputenc} 
\usepackage[T1]{fontenc}    
\usepackage{hyperref}       
\usepackage{url}            
\usepackage{booktabs}       
\usepackage{amsfonts}       
\usepackage{nicefrac}       
\usepackage{microtype}      
\usepackage{xcolor}         

\usepackage{amsmath}
\usepackage{amssymb}
\usepackage{mathtools}
\usepackage{amsthm}
\usepackage{multicol}
\usepackage{overpic}
\usepackage[capitalize,noabbrev]{cleveref}
\usepackage{mathrsfs}
\usepackage[ragged,
raggedright,
raggedleft
]{sidecap}   
\sidecaptionvpos{figure}{t,b} 

\usepackage{scalerel}
\usepackage{tikz}
\usetikzlibrary{svg.path}

\definecolor{orcidlogocol}{HTML}{A6CE39}
\tikzset{
  orcidlogo/.pic={
    \fill[orcidlogocol] svg{M256,128c0,70.7-57.3,128-128,128C57.3,256,0,198.7,0,128C0,57.3,57.3,0,128,0C198.7,0,256,57.3,256,128z};
    \fill[white] svg{M86.3,186.2H70.9V79.1h15.4v48.4V186.2z}
                 svg{M108.9,79.1h41.6c39.6,0,57,28.3,57,53.6c0,27.5-21.5,53.6-56.8,53.6h-41.8V79.1z M124.3,172.4h24.5c34.9,0,42.9-26.5,42.9-39.7c0-21.5-13.7-39.7-43.7-39.7h-23.7V172.4z}
                 svg{M88.7,56.8c0,5.5-4.5,10.1-10.1,10.1c-5.6,0-10.1-4.6-10.1-10.1c0-5.6,4.5-10.1,10.1-10.1C84.2,46.7,88.7,51.3,88.7,56.8z};
  }
}

\newcommand\orcidicon[1]{\href{https://orcid.org/#1}{\mbox{\scalerel*{
\begin{tikzpicture}[yscale=-1,transform shape]
\pic{orcidlogo};
\end{tikzpicture}
}{|}}}}

\newcommand*{\PP}{\text{P}}
\definecolor{orcidlogocol}{HTML}{A6CE39}

\title{Geometric path augmentation for inference  of sparsely observed stochastic nonlinear systems}

\author{%
  Dimitra Maoutsa\thanks{\href{https://dimitra-maoutsa.gitlab.io/}{https://dimitra-maoutsa.gitlab.io/} } \,\orcidicon{0000-0002-3553-8658} \\
  Technical University of Berlin\\
  Berlin, Germany \\
  \texttt{dimitra.maoutsa@tu-berlin.de} \\
}

\begin{document}

\maketitle
\vspace{-20pt}
\begin{abstract}
Stochastic evolution equations describing the dynamics of systems under the influence of both deterministic and stochastic forces are prevalent in all fields of science.
Yet, identifying these systems from sparse-in-time observations remains still a challenging endeavour.
Existing approaches focus either on the temporal structure of the observations by relying on conditional expectations, discarding thereby information ingrained in the geometry of the system's invariant density; or employ geometric approximations of the invariant density, which are nevertheless restricted to systems with conservative forces. 
Here we propose a method that reconciles these two paradigms. We introduce a new data-driven path augmentation scheme that takes the local observation geometry into account. By employing non-parametric inference on the augmented paths, we can efficiently identify the deterministic driving forces of the underlying system for systems observed at low sampling rates. 



\end{abstract}

\section{Introduction}
\label{sec:intro}
\vspace{-10pt}
Stochastic differential equations are particularly expressive dynamical models naturally fit for representing systems evolving on multiple time-scales~\cite{arnold2014phenotypic,lande1976natural,chandrasekhar1943stochastic,nelson2004biological}. Extracting stochastic evolution equations from such systems has been of major interest in most sciences~\cite{kuusela2004stochastic, sura2003stochastic,casadiego2018inferring,sura2003interpreting,frank2006stochastic,van2006deterministic, bergman2018inference,sarfati2017maximum,gottwald2017stochastic, maoutsa2022revealing,campioni2021inferring}. While identification of continuous time deterministic models has been largely resolved in the past~\cite{cremers1987construction,brunton2016discovering, daniels2015automated, mcgoff2015statistical}, 
the same is not true for their stochastic counterparts. Inference of stochastic systems is particularly challenging in settings
where observations are collected at low sampling rates (at large inter-observation intervals).

\begin{wrapfigure}{l}{0.35\textwidth}
\hspace{-20pt}
\includegraphics[width=0.40\textwidth]{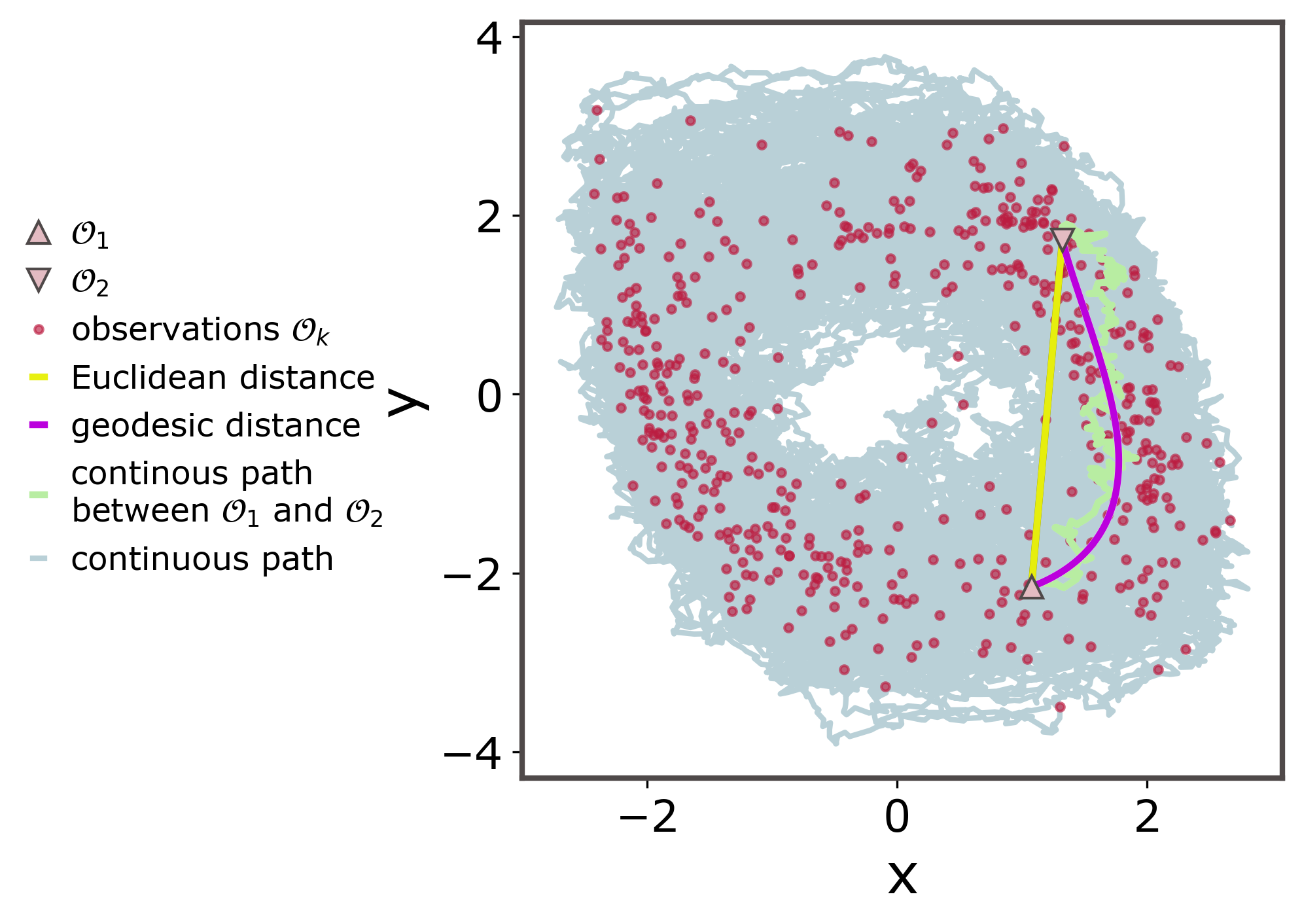}
  \caption{ \textbf{Considered state increments for low frequency observations under Gaussian likelihood assumptions.}  Euclidean distance (\emph{yellow line}) - used to compute the state increments between successive observations - does not account for the curvature of the invariant density. The geodesic curve (\emph{purple line}) provides a better approximation of the unobserved state of the system between successive observations (\emph{light green line}).}
  \label{fig:distances}
\end{wrapfigure}

Most of the existing methods for identifying the deterministic driving forces of stochastic systems rely either on approximations of the \textbf{invariant density} (e.g., density estimation~\cite{batz2016variational} or spectral methods like diffusion maps~\cite{singer2008non,nuske2021spectral,ionides2006inference,talmon2015intrinsic,dsilva2016data}) (\emph{\textbf{geometric methods}}), or consider the \textbf{temporal structure} of the observations by computing conditional expectations of state increments~\cite{friedrich1997description,siegert1998analysis,ragwitz2001indispensable,boninsegna2018sparse,lamouroux2009kernel,golightly2008bayesian,papaspiliopoulos2012nonparametric,sermaidis2013markov,beskos2006retrospective,chib2006likelihood,yildiz2018learning,friedrich2011approaching} (\emph{\textbf{temporal methods}}). 
However, geometric methods are limited only to systems with conservative forces by assuming either  
that the drift is the gradient of a potential~\cite{berry2018iterated,batz2016variational, talmon2015intrinsic}, or that state variables are completely decoupled~\cite{singer2008non}. On the other hand, temporal methods perform poorly in settings with large inter-observation intervals (\textbf{\emph{sparse observations}})~\cite{batz2018approximate}, since the state increments computed from the observations in those settings (see Appendix Eq.~\eqref{eq:increments}) do not reflect the actual underlying dynamics (Fig.~\ref{fig:distances}).

\begin{figure}
  \hspace{-80pt}
  \begin{overpic}[width=1.35\textwidth]{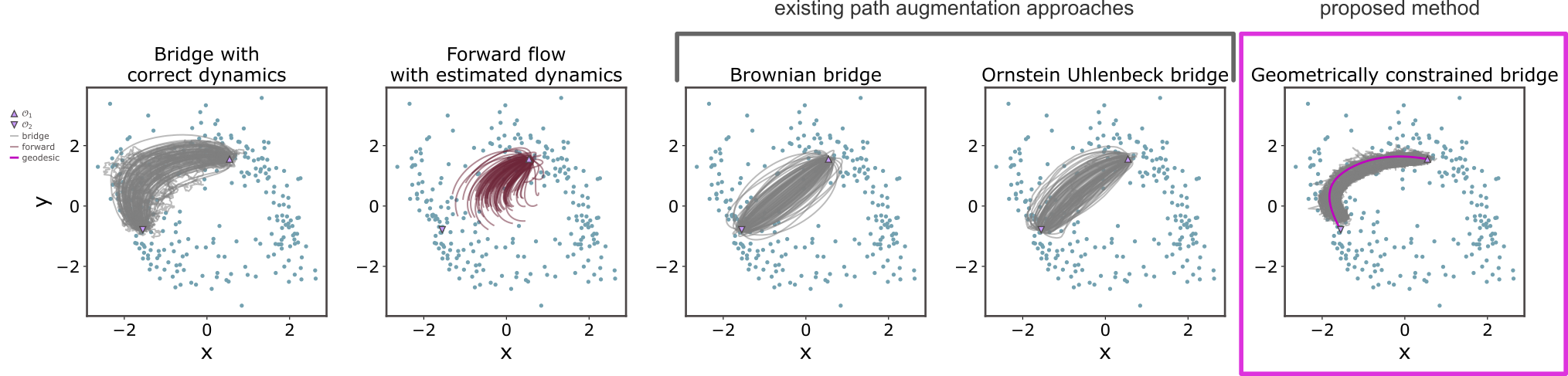}
  \put(3,25){a.}
  \put(23,25){b.}
  \put(43,25){c.}
  \put(63,25){d.}
  \put(80,25){e.}
  \end{overpic}
  \caption{ \textbf{Existing path augmentation strategies match poorly the underlying transition density between consecutive observations underestimating its curvature.} \textbf{a.)}  Stochastic bridge marginal density (\emph{grey}) between two successive observations $\mathcal{O}_1$ and $\mathcal{O}_2$ (\emph{pink triangles}) following the ground truth dynamics.
    \textbf{b.)}  Forward probability flow with estimated dynamics with a Gaussian likelihood (\emph{maroon}) matches poorly the correct transition density and often fails to reach the second observation $\mathcal{O}_2$ (\emph{downward pink triangle}).  Common path augmentation strategies employ either: \textbf{c.)} Brownian bridges, or \textbf{d.)} Ornstein Uhlenbeck (linear) bridge marginals resulting from local linearisations of the estimated drift with Gaussian likelihood. Both approaches match poorly the correct transition density, because they underestimate its curvature. \textbf{e.)} The proposed geometrically constrained path augmentation provides a better approximation of the underlying transition density by forcing the bridge paths towards the geodesic curve that connects consecutive observations on the manifold induced by the observations.     } 
  \label{fig:path_augmentations}
\end{figure}
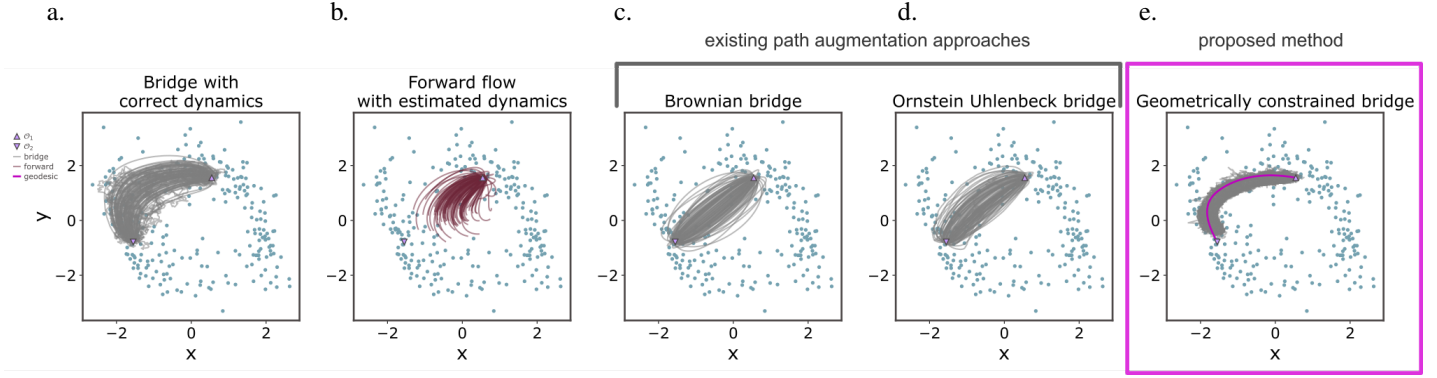

To mitigate the effect of sparse observations, a subset of the temporal methods employs \textbf{path augmentation}\footnote{Here we employ the term \textbf{path augmentation} to refer to what is widely known as \textbf{data augmentation}. We resorted to this term because we consider it as more elegant and better descriptive of the proposed augmentation process, while the term 'data augmentation' is considerably vague.} to approximate the transition densities between successive observations by sampling \textbf{diffusion bridges}, i.e., diffusion processes constrained by their initial and terminal state~\cite{golightly2008bayesian,papaspiliopoulos2012nonparametric,sermaidis2013markov,beskos2006retrospective,chib2006likelihood}. Yet, the majority of the \emph{non-parametric} approaches employ simplified bridge dynamics (e.g., Brownian~\cite{chib2006likelihood,golightly2008bayesian} or Ornstein-Uhlenbeck bridges~\cite{batz2018approximate}) that do not accurately reflect the underlying transition densities when the observed system is nonlinear (Fig.~\ref{fig:path_augmentations} c. and d.).

An alternative path augmentation strategy would consider a coarse drift estimate (e.g., by assuming a Gaussian likelihood between observations, see Eq.~\eqref{eq:SDE_likelihood}), and would subsequently employ a stochastic bridge sampler~\cite{maoutsa2022deterministic, maoutsa2021deterministica,debo2021simulating} to construct stochastic bridges with the estimated nonlinear drift. However, for large inter-observation intervals, the observations have zero probability under the law of the estimated diffusion (Fig.~\ref{fig:path_augmentations} b.). Consequently, any attempt to construct diffusion bridges between consecutive observations following the estimated dynamics will -- depending on the employed framework -- either show slow convergence rates, or fail altogether.

Here we propose an alternative approach. We postulate that the augmented paths should lie \textbf{in the vicinity} of the \textbf{geodesic curves} (Fig.~\ref{fig:distances}b. magenta) that connect consecutive observations on the manifold induced by the invariant density of the system. To that end we introduce a path augmentation framework that constructs \textbf{geometrically constrained bridges} \footnote{Formally these constructs are no longer diffusion bridges but constrained diffusion paths. Here we overextend the notion of diffusion bridges to contrast it against the commonly employed diffusion bridges for path augmentation.} by forcing the augmented paths towards the respective geodesics that connect consecutive observations (Fig.~\ref{fig:path_augmentations} e.). To that end we employ the stochastic control framework introduced in~\cite{maoutsa2021deterministica,maoutsa2022deterministic} with path constraints that guide the augmented paths towards the geodesic curves that connect successive observations.
\vspace{-10pt}
\section{Setting}
\vspace{-10pt}
We consider stochastic systems described by
stochastic differential equations (SDE) of the form
\begin{equation}\label{eq:system}
 \small \text{d}X_t = f(X_t) \text{d}t  + \sigma \,\text{d}\boldsymbol{\beta}_t, \qquad X_0 = x_0 \ ,
\end{equation}where $f(\cdot):R^d \rightarrow R^d$ denotes the deterministic driving forces (\emph{drift function}), and $\sigma \text{d}\boldsymbol{\beta}_t$ represents the random forces acting on the system (\emph{diffusion}). Here $\sigma \in \mathcal{R}^{d\times d}$ denotes the noise amplitude, and $\boldsymbol{\beta}$ the d-dimensional vector of independent Wiener processes. Onwards we consider Ito interpretation of stochastic integrals. 
We observe the system through an observation process $\mathcal{O}_k = \psi(X_{k \tau})$, where ${X}_{k\tau}\dot{=}{X}_t\vert{_{t=\tau k}}$, with $k =1, 2, \dots, K$ observations measuring the system state at \textbf{inter-observation intervals} $\tau$. For simplicity, we consider identity functions for the observation process, i.e., $\psi(x)=x$, but the formalism easily generalises for more general functions.

\paragraph{High-frequency observations.} For sufficiently fine observation timegrids, we assume that observations represent the system state in continuous time, i.e., that we observe the continuous path $X_{0:T}$. Thereby we can estimate the drift by approximating the first order Kramers-Moyal coefficient~\cite{tabar2019kramers} through empirically estimating conditional expectations of state increments~\cite{friedrich2000extracting,ragwitz2001indispensable, boninsegna2018sparse,siegert1998analysis}. Analogous Bayesian non-parametric methods~\cite{ruttor2013approximate} consider that the transition probabilities between observations are Gaussian for $\text{d}t\rightarrow 0$, resulting in a (Gaussian) likelihood for the observations (see Sec.~\ref{appsec:b} Eq.~\ref{apeq:SDE_likelihood})
\begin{equation} \label{eq:SDE_likelihood}
\small    \mathcal{L}(X_{0:T}\mid f) = \exp \left[ -\frac{1}{2} \int^T_0 \| f(X_t)\|_{\sigma^2}^2  \text{d}t + \int^T_0 \langle f(X_t), X_{t+\text{d}t}-X_t \rangle \text{d}t \right] ,
\end{equation}
and impose a Gaussian process prior on the function values $f$ (Eq.~\eqref{eq7:full_gp_mean}). In Eq.~\eqref{eq:SDE_likelihood} we introduced the notation ${\langle u, v\rangle \dot{=} u^{\top} \cdot \sigma^{-2} v}$ and $\|u \|_{\sigma^2}  \dot{=} u^{\top} \cdot \sigma^{-2} u$.
\vspace{-10pt}
\paragraph{Low-frequency observations.} As the inter-observation interval $\tau$ increases, the Gaussian likelihood (Eq.~\eqref{eq:SDE_likelihood}) assumed between two successive observations is no longer valid if Eq.~\eqref{eq:system} is non-linear. Similarly, the state increments $X_{t+\tau} - X_t$ computed in this setting do not accurately represent the underlying dynamics (Fig.~\ref{fig:distances} a.).
The likelihood for the drift $ \PP(\{\mathcal{O}_k\}_{k=1}^{K}|f)$ for such settings takes the form of a \emph{path integral}
\begin{equation}\label{eq7:path_integral_likelihood}
 \small   \PP(\{\mathcal{O}_k\}_{k=1}^{K}\mid f)= \int \PP (\{\mathcal{O}_k\}_{k=1}^{K}, X_{0:T}\mid f) \mathcal{D}(X_{0:T}) = \int \PP(\{\mathcal{O}_k\}_{k=1}^{K}\mid X_{0:T}) \PP(X_{0:T}|f) \mathcal{D}(X_{0:T}),
\end{equation}
where $\{\mathcal{O}_k\}_{k=1}^{K}$ denotes the set of $K$ discrete time observations, 
$\PP(X_{0:T}|f)$ the prior path probability resulting from the system of Eq.~\eqref{eq:system}, $\mathcal{D}(X_{0:T}) $ identifies the formal volume element on the path space, while $\PP(\{\mathcal{O}_k\}_{k=1}^{K}| X_{0:T})$ stands for the likelihood of observations given the latent path $X_{0:T}$.

From a geometric perspective, we can consider that the invariant density of the system can be approximated with a (low dimensional) manifold induced by the nonlinear system dynamics. The observations are essentially samples of that manifold. For low-frequency observations, Euclidean distances employed for computing the state increments $X_{t+\tau}-X_t$ do not consider the geometry induced by the nonlinear dynamics, and thereby underestimate the curvature of the transition density between consecutive observations (Figure~\ref{fig:distances}).
\vspace{-10pt}
\section{Method}
\vspace{-10pt}
Since the likelihood of Eq.~\eqref{eq7:path_integral_likelihood} is intractable, we consider the unobserved continuous path as latent random variables $X_{0:T}$, and employ Expectation Maximisation (EM)~\cite{dempster1977maximum} to identify a maximum a posteriori estimate for the drift function. Similar parametric~\cite{elerian2001likelihood,sermaidis2013markov} and non-parametric~\cite{batz2018approximate,ruttor2013approximate} methods have addressed the drift inference in the past, targeting mainly high-frequency observation settings. Our approach here is inspired by the non-parametric method followed in~\cite{batz2018approximate,ruttor2013approximate} with two key innovations:
\begin{itemize}
    \item[\textbf{\textcolor{mymaroon2}{(i)}}]  We employ a path augmentation scheme following the \textbf{estimated nonlinear dynamics} resulting from inference with the Gaussian likelihood of Eq.~\eqref{eq:SDE_likelihood} (as opposed to local linear approximations of these dynamics proposed in~\cite{batz2018approximate}).
    \item[\textbf{\textcolor{mymaroon2}{(ii)}}]  Importantly, we further \textbf{constrain the augmented paths to match the} \textbf{geometry of the invariant density} between consecutive observations (Fig.~\ref{fig:distances} b.).
\end{itemize}

  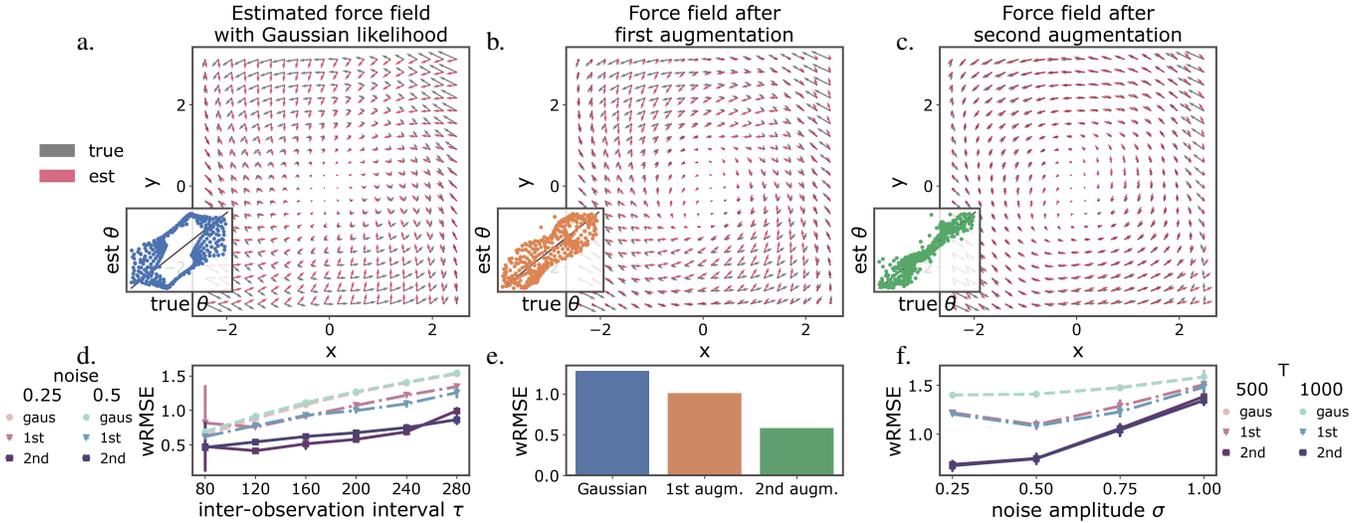
\begin{figure}
\hspace{-65pt}
  \begin{overpic}[width=1.3\textwidth]{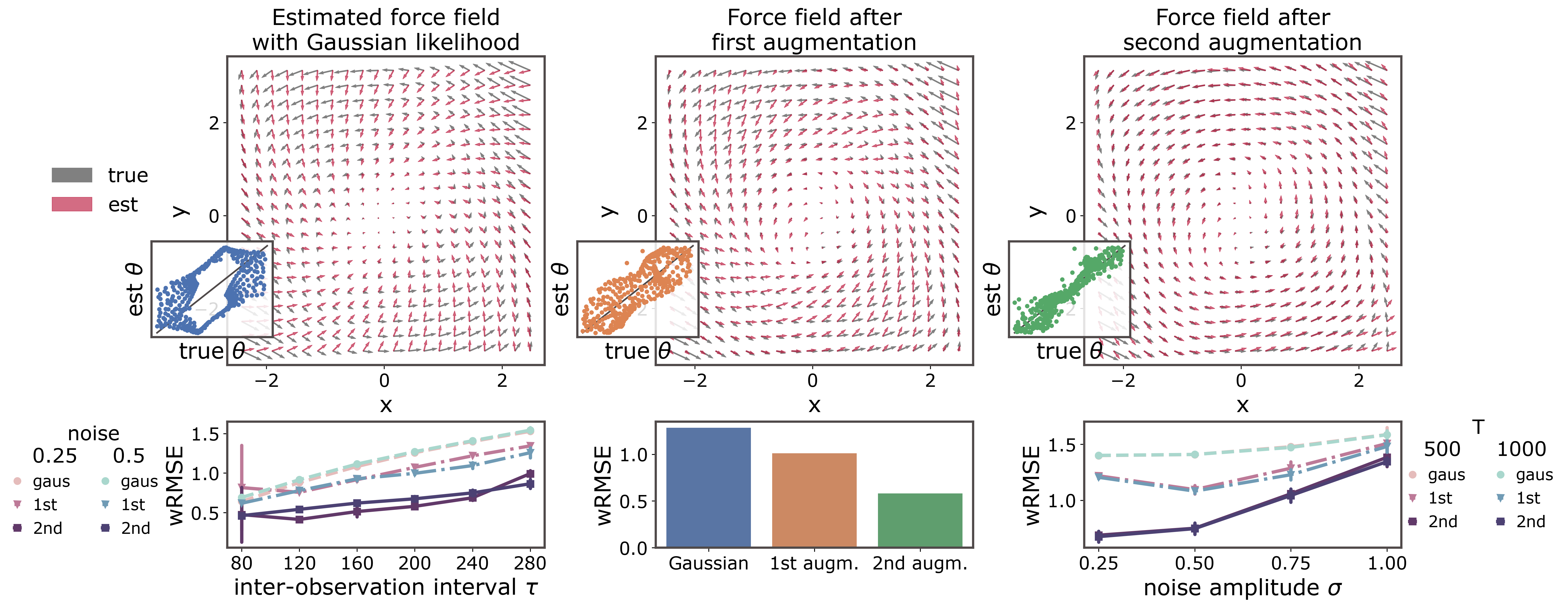}
  \put(6,35){a.}
  \put(36,35){b.}
  \put(66,35){c.}
  \put(6,12){d.}
  \put(36,12){e.}
  \put(66,12){f.}
  \end{overpic}
  \caption{ \textbf{Proposed path augmentation after two iterations already provides a good approximation of underlying drift.} Estimated (\emph{red}) and true (\emph{grey}) force field with \textbf{a.)} Gaussian likelihood   
    \textbf{b.)} after one and \textbf{c.)} after second iteration of augmentations. (\textbf{insets}) Ground truth against estimated angles for each point on the two dimensional grid. \textbf{e.)} Weighted root mean square error (wRMSE) for estimated drifts after each iteration for the presented example. The weights for averaging the error at each grid point are obtained from a kernel density estimation on the observations $\{\mathcal{O}_k\}^K_{k=1}$. \textbf{d.)} wRMSE against inter-observation interval $\tau$ for different noise conditions $\sigma=\{0.25,0.5\}$ for drift estimated with a Gaussian likelihood (\emph{gaus}-circles), after first augmentation (\emph{1st}-triangles), and after second augmentation (\emph{2nd}-squares) for $T=500$. \textbf{f.)} wRMSE against noise amplitude $\sigma$ in the system for different trajectory durations $T=\{500,1000\}$ time units for inter-observation interval $\tau=240$. Markers follow the same coding as in d.). Errorbars indicate one standard deviation over $5$ independent realisations.    } 
  \label{fig:res} \vspace{-6pt}
\end{figure}

\vspace{-8pt}

We follow an iterative algorithm, where at each iteration $n$ we perform the two following steps:\\
\textbf{(1.)}
     An \textbf{E(xpectation) step}, where given a drift estimate $\hat{f}^n$ we construct an approximate posterior over the latent variables $Q(X_{0:T}) \approx \PP(X_{0:T}|\{\mathcal{O}\}^K_{k=1}, \hat{f}^n(x))$.\\
\textbf{(2.)}  
     A \textbf{M(aximisation) step}, where we update the drift estimation.
     \vspace{-10pt}
\paragraph{$\bullet$ Approximate posterior over paths. (E-step)}
We approximate the continuous path trajectory $X_{0:T}$ between observations by a posterior path measure defined as the minimiser of the free energy \vspace{-6pt}
\begin{equation} \label{eq:free_energy2}
 \small   \mathcal{F}[Q] = \frac{1}{2} \int \limits^T_0 \int \Big[ \|g(x,t) - \hat{f}(x)\|_{\sigma^2}^2 + {U_{\mathcal{O}}(x,t)} +{U_{\mathcal{G}}(x,t)}   \Big] \, q_t(x)\, \text{d}x \,\text{d}t.
\end{equation}
The term ${U_{\mathcal{O}}(x,t) \dot{=} - \sum_{t_k} \ln \PP(\mathcal{O}_k|x) \delta(t- t_k)}$ forces the latent path to pass through the observations (or close to them depending on the observation process), while ${U_{\mathcal{G}}(x,t) \dot{=} \| \Gamma_t -x \|^2 }$ guides the latent path towards the geodesic curves $\gamma^k_{t'}$ that connect consecutive observations on the manifold $\mathcal{M}$ induced by the system's invariant density (Sec.~\ref{appsec:with}). Here we denote $\Gamma_t\dot{=} \{\gamma^k_{t'}\}_{t=(k-1)\tau+t' \tau}$, where $\gamma^k_{t'}$ is the geodesic connecting $\mathcal{O}_k$ and $\mathcal{O}_{k+1}$, and $t'\in [0,1]$. We identify the geodesic $\gamma^k_{t'}$ for each interval by learning the local metric of the manifold $\mathcal{M}$ (see Sec.~\ref{appsec:with} and~\cite{arvanitidis2019fast}). 

Following~\cite{opper2019variational}, for each inter-observation interval $[\mathcal{O}_k, \mathcal{O}_{k+1}]$ we identify the posterior path measure (minimiser of Eq.~\eqref{eq:free_energy2}) by the solution of a stochastic optimal control problem~\cite{maoutsa2022deterministic,maoutsa2021deterministica, maoutsa2022revealing} with the objective to obtain a time-dependent drift adjustment $u(x,t):=g(x,t) - \hat{f}(x)$ for the system with drift $\hat{f}(x)$ with {\textbf{initial and terminal constraints}} determined by $U_{\mathcal{O}}(x,t)$, and additional {\textbf{path constraints}} $U_{\mathcal{G}}(x,t)$.
\vspace{-10pt}
\paragraph{$\bullet$ Drift estimation. (M-step)} To estimate the drift from a sampled latent path, we assume a Gaussian process prior over function values and employ a sparse kernel approximation similar to~\cite{batz2018approximate} (see Sec.~\ref{appsec:drift_inference} for details).
\vspace{-10pt}

\section{Numerical experiments}\vspace{-5pt}
To demonstrate the performance of the proposed method we performed systematic estimations for a two-dimensional
Van der Pol oscillator under different noise conditions $\sigma$, observed at different inter-observation intervals $\tau$ for different lengths of trajectories $T$ (see Sec.~\ref{sec:details}). For the examined noise amplitudes (Fig.~\ref{fig:res} f.) and for inter-observation intervals that result in more than one observation per oscillation period (Fig.~\ref{fig:res}d.), the proposed path augmentation algorithm improves the naive estimation with Gaussian assumptions within two iterations for most noise amplitudes(Fig.~\ref{fig:res}). For increasing noise the improvement contributed by our approach decreases (Fig.~\ref{fig:res}f.), but is nevertheless not negligible.
\vspace{-10pt}
\section{Conclusion and Discussion}\vspace{-5pt}
We introduced a new method for identifying stochastic systems from sparse-in-time observations of the system's state. We proposed a path augmentation strategy that employs the nonlinear dynamics of a coarse drift estimate, and further constrains the augmented paths to follow the local geometry of the system's invariant density. We found that the proposed approach provides efficient recovery of the underlying drift function for periodic or quasi-periodic systems under several noise conditions.

\vspace{-10pt}
\paragraph{Geometric constraints for inference.} Our method reconciles approaches that rely purely on the temporal structure of the observations with those that approximate the invariant density and ignore the temporal order of measurements. With the recent development of the field of geometric statistics~\cite{miolane2020geomstats, sommer2020probabilistic}, and the surge of interest on the concept of manifold hypothesis~\cite{fefferman2016testing,shnitzer2020manifold}, i.e., the consideration that often the state of multi-dimensional dynamical systems is confined on low dimensional regions of the state space,
several inference methods have tried to merge geometric and temporal perspectives for identification of stochastic systems. In the \emph{Langevin regression} framework~\cite{callaham2021nonlinear}, Callaham \emph{et al.} compute the Kramers-Moyal coefficients, and account for misestimation due to low sampling rate by solving the adjoint Fokker-Planck equation for the coefficients as proposed by Lade~\cite{lade2009finite}. They incorporate geometric constraints by additionally regularising by moment matching between the observation density and the stationary Fokker-Planck probability density of the estimated SDE model. In~\cite{tong2020trajectorynet} Tong \emph{et al.} consider the manifold of the observations for inference of cellular dynamics. Their method employs dynamic optimal transport to interpolate between measured distributions constrained to lie in the vicinity of the observations. This approach has the same intuitions with our method, however Tong et al. do not employ stochastic differential equations to model the inherently stochastic cellular dynamics. Moreover they do not attempt any modeling of the underlying geometry of the data, but consider only constraints that penalise distances to individual observations. Shnitzer et al.~\cite{shnitzer2020manifold, shnitzer2016manifold}
employ diffusion maps to approximate the eigenfunctions of the backward Kolmogorov operator (the generator of the stochastic Koopman operator~\cite{giannakis2019data,vcrnjaric2020koopman}), and - since the eigenfunctions follow linear evolution equations - they evolve the dominant operator eigenspectrum with a Kalman filter to account for the temporal order of the observations.
  However their approach is limited to conservative systems, and assumes the existence of a spectral gap on the spectrum of the approximated operator, excluding thereby systems with continuous spectra, e.g.  chaotic systems~\cite{koopman1932dynamical,mezic2005spectral}.
\vspace{-10pt}
\paragraph{Geodesic curves and the most probable path in the Onsager-Machlup sense for stochastic processes.} 
The theoretical underpinnings of our work can be traced back to the work of Onsager and Machlup~\cite{onsager1953fluctuations} and the computation of the \textbf{most probable path} (\textbf{MPP}) of a diffusion process between two predetermined states. 
Earlier work has employed the \textbf{Onsager-Machlup (OM) function} as Lagrangian to derive an expression for the \textbf{MPP} in terms of state variables and the drift of the diffusion process~\cite{fujita1982onsager,takahashi1981probability,durr1978onsager,graham1977path,ito1978probabilistic,stratonovich1971probability,sommer2015anisotropic}. The resulting Lagrangian involves the energy of the path (see Sec.~\ref{app:OM}), which is the same objective used to identify geodesics (see Sec.~\ref{appsec:with}). In our framework, to identify the geodesics between successive observations we assumed as smooth manifold the $\mathcal{R}^d$ with associated Riemannian metric $h$ learned from the data. However the underlying SDE is defined on $\mathcal{R}^d$ under the Euclidean metric~\cite{capitaine2000onsager}. Different metrics in the definition of diffusion processes result in different generators, and thus in different path probabilities for each process. It would be an interesting theoretical result to calculate the transformation induced by the changing the Riemannian metric in the definition of the process. To the best of our knowledge such a result is not available in the literature, but is also not trivial. The change of metric induces a change in the diffusivity of the process, so a direct Girsanov transformation is not feasible. Yet, 
it may be possible to employ a transformation akin to an inverse Lamperti transformation~\cite{oksendal2003stochastic} to express the drift of the process in terms of a diffusion with multiplicative noise that would have induced the change in metric learned from the observations (see Sec.~\ref{app:OM}). 
 Finally, the connection to the \textbf{OM} functional already hints to an alternative method to obtain an estimate for the unknown drift with geometric considerations that obviates the computationally costly simulation of continuous paths. 

\vspace{-10pt}
\paragraph{Limitations.}
Our approach is limited to systems where the invariant density can be approximated by a manifold on which one can identify geodesics. Additionally by construction the method implicitly assumes that the invariant system's density is approximately uniformly sampled. However we foresee that by employing more advanced tools from geometric statistics~\cite{miolane2020geomstats, sommer2020probabilistic} the framework may be applicable for non-uniformly sampled invariant densities. Our experiments have shown that the proposed approach is better suited for systems with \emph{cyclic balance}~\cite{tomita1974irreversible}, i.e., with stationary fluctuating probability currents in the steady state. Systems with non-fluctuating currents in the stationary state are nevertheless effectively recovered with existing methods that rely on assumptions of conservative forces.


\section*{Acknowledgements}
We are indebted to Nina Miolane for providing detailed information on how to compute geodesics from manifolds approximated with Variational Autoencoders. We further thank Georgios Arvanitidis for maintaining a publicly available repository with the algorithm employed here to construct geodesics, Stefan Sommer for answering questions on transformations of diffusion processes on manifolds, and Prof. Manfred Opper for prompting us to work on this problem and for providing initial guidance. We further acknowledge that previous work from the Python~\cite{van1995python}, numpy~\cite{harris2020array}, scipy~\cite{2020SciPy-NMeth}, matplotlib~\cite{Hunter:2007}, seaborn~\cite{waskom2021seaborn}, GPflow~\cite{GPflow2017}, pyEMD~\cite{pele2009}, and pytorch~\cite{paszke2017automatic} communities facilitated the implementation of the computational part of this work.

An implementation of this work will be released in the following repository: \href{https://github.com/dimitra-maoutsa/Geometric-path-augmentation-for-SDEs}{https://github.com/dimitra-maoutsa/Geometric-path-augmentation-for-SDEs} once the article gets submitted for archival publication.

\section*{Broader Impact Statement}

We introduced a new path augmentation method that allows for efficient inference of stochastic systems observed at large inter-observation intervals. Our contribution aims to 
highlight the need for incorporating notions from the rapidly developing field of geometric statistics into the area of model discovery of stochastic systems.  While geometric and topological properties of invariant densities for deterministic systems have been exceedingly studied in the past, the same is not true for their stochastic counterparts, and in particular of systems described by stochastic differential equations. 

Our work aims further to highlight that data augmentation frameworks in settings where the amount of augmented data dominates the number of observations may lead to more accurate inferences by incorporating domain knowledge or other type of information in the augmentation (like here information regarding the geometry of the system's invariant density). Many of the algorithms employed with data augmentation frameworks exhibit only \textbf{local convergence}, e.g., the Expectation Maximisation algorithm employed here~\cite{romero2019convergence}. In settings where the initial estimate strongly deviates from its true value, naive data augmentation strategies might therefore converge to sub-optimal solutions, that do not reflect the ground truth.

We do not foresee any direct social impact of our work. However we acknowledge that stochastic systems may be used for military purposes and financial engineering, however the proposed method does not directly propose interventions to the observed system that may lead to unfavourable consequences.

Diffusive systems are prevalent in several scientific fields, such as parts of physics, biology, neuroscience, and ecology.
We foresee that this work may benefit these disciplines by providing a tool for identifying systems of interest.

\printbibliography

\section*{Checklist}


\begin{enumerate}

\item For all authors...
\begin{enumerate}
  \item Do the main claims made in the abstract and introduction accurately reflect the paper's contributions and scope?
    \answerYes{We provide numerical results to justify our claims.}
  \item Did you describe the limitations of your work?
    \answerYes{}
  \item Did you discuss any potential negative societal impacts of your work?
    \answerYes{In the impact statement.}
  \item Have you read the ethics review guidelines and ensured that your paper conforms to them?
    \answerYes{}
\end{enumerate}

\item If you are including theoretical results...
\begin{enumerate}
  \item Did you state the full set of assumptions of all theoretical results?
    \answerNA{}
        \item Did you include complete proofs of all theoretical results?
    \answerNA{}
\end{enumerate}

\item If you ran experiments...
\begin{enumerate}
  \item Did you include the code, data, and instructions needed to reproduce the main experimental results (either in the supplemental material or as a URL)?
    \answerNo{Not yet. The work will be submitted for archival publication, therefore for now we do not open source our code. However, our work combines already published frameworks~\cite{arvanitidis2019fast,maoutsa2022deterministic, batz2018approximate}. The articles~\cite{arvanitidis2019fast,maoutsa2022deterministic} provide github repositories with available implementations, which are the ones we used. For~\cite{batz2018approximate} we re-implemented the drift inference as described in the main paper and the supplement. For archival reasons, and to make our code discoverable for people who will read the camera-ready version of the article in the future, we will provide a link to a non-populated repository, that will host the implementation once our full article is ready for submission.}
  \item Did you specify all the training details (e.g., data splits, hyperparameters, how they were chosen)?
    \answerYes{}
        \item Did you report error bars (e.g., with respect to the random seed after running experiments multiple times)?
    \answerNA{}
        \item Did you include the total amount of compute and the type of resources used (e.g., type of GPUs, internal cluster, or cloud provider)?
    \answerYes{One run of our framework for a two dimensional system requires around 4-6 hours (depending on the amount of data) on a laptop with Intel Core i7@ 1.80GHz CPU.}
\end{enumerate}

\item If you are using existing assets (e.g., code, data, models) or curating/releasing new assets...
\begin{enumerate}
  \item If your work uses existing assets, did you cite the creators?
    \answerYes{}
  \item Did you mention the license of the assets?
    \answerNA{}
  \item Did you include any new assets either in the supplemental material or as a URL?
    \answerNo{Not yet. We will release the framework once the paper will be submitted for archival publication.}
  \item Did you discuss whether and how consent was obtained from people whose data you're using/curating?
    \answerNA{}
  \item Did you discuss whether the data you are using/curating contains personally identifiable information or offensive content?
    \answerNA{}
\end{enumerate}

\item If you used crowdsourcing or conducted research with human subjects...
\begin{enumerate}
  \item Did you include the full text of instructions given to participants and screenshots, if applicable?
    \answerNA{}
  \item Did you describe any potential participant risks, with links to Institutional Review Board (IRB) approvals, if applicable?
    \answerNA{}
  \item Did you include the estimated hourly wage paid to participants and the total amount spent on participant compensation?
    \answerNA{}
\end{enumerate}

\end{enumerate}

\newpage
\appendix
\onecolumn
\appendixpage

\startcontents[sections]
\printcontents[sections]{l}{1}{\setcounter{tocdepth}{3}}

\section{Drift inference for high and low frequency observations}\label{appsec:b}
We consider systems whose evolution is captured by the stochastic differential equation Eq.~\eqref{eq:system}.
\paragraph{High frequency observations.} When the system path $X_{0:T}$ is observed in continuous time, the infinitesimal transition probabilities of the diffusion process between consecutive observations are Gaussian, i.e.,
\begin{equation}
    \PP_f(X_{0:T} \mid f) \propto \exp \left( -\frac{1}{2 \text{d}t} \sum_t  \| X_{t+\text{d}t} -X_t - f(X_t)\text{d}t\|_{\sigma^2}^2     \right).
\end{equation}
In turn, the transition probability of (discretised) Wiener paths $P_{\mathcal{W}}(X_{0:T}) $ (i.e., paths from a drift-less process) can be expressed as
\begin{equation}
    \PP_{\mathcal{W}}(X_{0:T})=\exp \left( -\frac{1}{2 \text{d}t} \sum_t  \| X_{t+\text{d}t} -X_t\|_{\sigma^2}^2     \right),
\end{equation}
where $\|u \|_{\sigma^2}  \dot{=} u^{\top} \cdot {\sigma}^{-2} u$ denotes the weighted norm with $D\dot{=}\sigma^2$ indicating the noise covariance. 
We can thus express the likelihood for the drift $f$ by the Radon-Nykodym derivative between $P_f(X_{0:T}|f)$ and $\PP_{\mathcal{W}}(X_{0:T})$
 for paths $X_{0:T}$ within the time interval $[0,\,T]$
~\cite{liptser2013statistics}
\begin{equation} \label{apeq:SDE_likelihood}
    \mathcal{L}(X_{0:T} \mid f) = \exp \left[ -\frac{1}{2} \sum_t \| f(X_t)\|_{\sigma^2}^2  \text{d}t + \sum_t \langle f(X_t), X_{t+\text{d}t}-X_t \rangle_{\sigma^2}  \right],\end{equation}
where for brevity we have introduced the notation $\langle u, v\rangle \dot{=} u^{\top} \cdot {\sigma}^{-2} v$ for the weighted inner product with respect to the inverse noise covariance ${\sigma}^{-2}$. This expression results from applying the Girsanov theorem on the path measures induced by a process with drift $f$ and a Wiener process, with same diffusion $\sigma$, and employing an Euler-Maruyama discretisation on the continuous path $X_{0:T}$.

The likelihood of a continuously observed path of the SDE (Eq.~\eqref{apeq:SDE_likelihood}) has a quadratic form in terms of the drift function. Therefore a Gaussian measure over function values (Gaussian process) is a natural conjugate prior for this likelihood. To identify the drift in a non-parametric form, we assume a Gaussian process prior for the function values $f \sim \PP_0({f}) =\text{GP}(m^f, k^f)$, where $m^f$ and $k^f$ denote the mean and covariance function of the Gaussian process ~\cite{ruttor2013approximate}. The prior measure can be written as
\begin{equation}
    \PP_0({f})  = \exp\left[-\frac{1}{2} \int \int f(x) \left(k^f(x,x') \right)^{-1}f(x') \text{d}x \text{d}x'\right],
\end{equation}
if we consider a zero mean Gaussian process $m^f=0$.

Bayesian inference for the drift function $f$ requires the computation of a probability distribution in the function space, the posterior probability distribution $\PP_f(f \mid X_{0:T})$. From the Bayes' rule the posterior can be expressed as 
\begin{equation}
    \PP_f(f\mid X_{0:T}) =\frac{ \PP_0(f) \mathcal{L}(X_{0:T} \mid f)}{Z} \propto \PP_0(f) \mathcal{L}(X_{0:T} \mid f),
\end{equation}
where $Z$ denotes a normalising factor defined as a path integral 
\begin{equation}
    Z = \int \PP_0(f) \mathcal{L}(X_{0:T} \mid f) \mathcal{D}f ,
\end{equation}
where $ \mathcal{D}f$ denotes integration over the Hilbert space $f: H_0[f] < \infty$ . Here we have expressed the prior probability over functions as $\PP_0(f) = e^{-H_0[f]}$.
In~\cite{ruttor2013approximate} the authors show that in the continuous time limit, nonparametric estimation of drift functions becomes equivalent to Gaussian process regression,  with the objective to identify the mapping from the system state $X_t$ to state increments $\text{d}X_t$~\cite{rasmussen2003gaussian}. More precisely, we consider $N$ observations of the system state $X_t$ as the regressor, with associated response variables 
\begin{equation}\label{eq:increments}
    Y_t = \frac{X_{t+\text{d}t}-X_t}{\text{d}t},
\end{equation} and denote the kernel function of the Gaussian process by $k(x,x')$. 

If we denote with $\mathcal{X}= \{X_t\}^{T-\text{d}t}_{t=0}$ and $\mathcal{Y}= \{Y_t\}^{T-\text{d}t}_{t=0}$ the set of state observations and observation increments, the mean of the posterior process over drift functions $f$ can be expressed as
\begin{equation}\label{eq7:full_gp_mean}
    \bar{f}(x) = k^f(x,\mathcal{X})^{\top} \left( \mathcal{K} + \frac{{\sigma}^{2}}{\text{d}t} I_N\right)^{-1} \mathcal{Y},
\end{equation}
where we abused the notation and denoted with $k^f(x, \mathcal{X})$ the vector resulting from evaluating the kernel $k^f$ at points $x$ and $\{\mathcal{O}_t\}^{K-1}_{k=1}$. Similarly $\mathcal{K} = k^f(\mathcal{X}, \mathcal{X})$ stands for the $(K-1)\times (K-1)$ matrix resulting from evaluation of the kernel on all observation pairs. 
In a similar vein, the posterior variance can be written as
\begin{equation} \label{eq7:full_gp_cov}
    \Sigma^2(x) = k^f(x,x) - k^f(x,\mathcal{X})^{\top} \left(\mathcal{K} + \frac{{\sigma}^{2}}{\text{d}t} \right)^{-1} k^f(x,\mathcal{X}),
\end{equation}
where the term ${\sigma}^{2}/\text{d}t$ plays the role of observation noise.

\paragraph{Low frequency observations.} When the inter-observation interval becomes large (\emph{low frequency observations}), the Gaussian likelihood of Eq.~\eqref{apeq:SDE_likelihood} becomes invalid, since for large inter-observation intervals the transition density is no longer Gaussian. Thus, drift estimation with Gaussian assumptions ~\cite{friedrich1997description,ruttor2013approximate} becomes inaccurate. To mitigate this issue Lade~\cite{lade2009finite} introduced a method to compute finite time corrections for the drift estimates, which has been applied (to the best of our knowledge) mostly to one dimensional problems~\cite{honisch2011estimation}. On the other hand, the statistics community has proposed path augmentation schemes that augment the observed trajectory to a nearly continuous-time trajectory by sampling a simplified system's dynamics between observations~\cite{golightly2008bayesian,papaspiliopoulos2012nonparametric,sermaidis2013markov,beskos2006retrospective,chib2006likelihood}.
However for large inter-observation intervals and for nonlinear systems the simplified dynamics employed for path augmentation match poorly the underlying path statistics, and these methods show poor convergence rates or fail to identify the correct dynamics (Figure~\ref{fig:path_augmentations} c. and d.). We point out here, that path augmentation with Ornstein Uhlenbeck bridges using as drift the local linearisation of the \textbf{correct} dynamics, provides a good approximation of the underlying transition density. However, during inference, the true underlying dynamics are unknown, and the proposed local linearisations on inaccurate drift estimates~\cite{batz2018approximate} perform poorly for low frequency observations.

Notice that as the inter-observation interval $\tau$ increases, the Gaussian likelihood assumed between two successive observations is no longer valid if the system is non-linear or when the noise is state dependent. 
The likelihood for the drift for such settings can be expressed in terms of a \emph{path integral}
\begin{equation}\label{apeq7:path_integral_likelihood}
    \PP(\mathcal{O}_{1:K}\mid f) = \int \PP(\mathcal{O}_{1:K}\mid X_{0:T}) \PP(X_{0:T}\mid f) \mathcal{D}(X_{0:T}),
\end{equation}
where $\mathcal{O}_{1:K}\dot{=}\{\mathcal{O}_k\}_{k=1}^{K}$ denotes the set of $K$ discrete time observations, 
$\PP(X_{0:T}\mid f)$ the prior path probability resulting from a diffusion process with drift $f(x)$, $\mathcal{D}(X_{0:T}) $ identifies the formal volume element on the path space, and $\PP(\mathcal{O}_{1:K}\mid X_{0:T})$ stands for the likelihood of observations given the latent path $X_{0:T}$.

However, the path integral of Eq.~\eqref{apeq7:path_integral_likelihood} is intractable for nonlinear systems, thus we need to simultaneously estimate the drift and latent state of the diffusion process, i.e., to approximate the joint posterior measure of latent paths and drift functions $\PP(X_{0:T},f \mid \mathcal{O}_{1:K})$. Therefore we consider the unobserved continuous path $X_{0:T}$ as latent random variables and employ an Expectation Maximisation (EM) algorithm to identify a maximum a posteriori estimate for the drift function. 
More precisely, we follow an iterative algorithm, where at each iteration $n$ we alternate between the two following steps:
\vspace{-5pt}
\begin{multicols}{2}
\begin{minipage}[c]{0.95\linewidth}
     An \textbf{Expectation} step, where given a drift estimate $\hat{f}^n(x)$ we construct an approximate posterior over the latent variables ${Q(X_{0:T}) \approx \PP(X_{0:T}\mid \mathcal{O}_{1:K}, \hat{f}^n(x))}$, and compute the expected log-likelihood of the augmented path 
    \begin{equation} \label{eq:estep}
    \mathfrak{L}\big(\hat{f}^n(x), Q\big) = \mathbb{E}_Q\Big[\ln \mathscr{L}\big(X_{0:T}\mid \hat{f}^n(x)\big) \Big].
    \end{equation}
    \end{minipage}
    
     A \textbf{Maximisation} step, where we update the drift estimation by maximising the expected log likelihood
    \begin{equation} \label{eq:mstep}
        f^{n+1}(x) = \arg \max_f \Big[\mathfrak{L}\big(f^n(x),Q\big)-\ln \PP_0\big(f^n(x)\big) \Big].
    \end{equation}
\end{multicols}
In Eq.~\eqref{eq:mstep} $\PP_0$ denotes the Gaussian process prior over function values.

\subsection{Approximate posterior over paths.} 

Here we first formulate the approximate posterior over paths (conditional distribution for the path given the observations) by considering only individual observations as constraints (Section~\ref{appsec:without}). However, this approach results computationally taxing calculations during path augmentation, since the observations are atypical states of the initially estimated drift. To overcome this issue, we subsequently extend the formalism (Section~\ref{appsec:with}) to incorporate constraints that consider also the local geometry of the observations.

\subsubsection{Approximate posterior over paths \underline{without} geometric constraints.}\label{appsec:without}
Given a drift function (or a drift estimate) $\hat{f}(x)$ we can apply variational techniques to approximate the posterior measure over the latent path conditioned on the observations $\mathcal{O}_{1:K}$. We consider that the prior process (the process without considering the observations $\mathcal{O}_{1:K}$) is described by the equation
\begin{equation} \label{appeq:prior_sde}
  \PP(X_{0:T}\mid \hat{f}): \qquad  \text{d}X_t = \hat{f}(X_t) \text{d}t + \sigma \text{d}\boldsymbol{\beta}_t.
\end{equation}
We will define an approximating (posterior) process that is conditioned on the observations. The conditioned process is also a diffusion process with the same diffusion as Eq.~\eqref{appeq:prior_sde} but with a modified, time-dependent drift $g(x,t)$ that accounts for the observations~\cite{chetrite2015variational,majumdar2015effective}. 
We identify the approximate posterior measure $Q$ with the posterior measure induced by an approximating process that is conditioned by the observations $\mathcal{O}_{1:K}$~\cite{opper2019variational},
with governing equation
\begin{equation} \label{appeq:sde_q}
    Q(X_{0:T}): \qquad \text{d}X_t = g(X_t,t) \text{d}t + \sigma \text{d}\boldsymbol{\beta}_t=\left(\hat{f}(X_t) + \sigma^2 u(X_t,t) \right) \text{d}t +  \sigma \text{d}\boldsymbol{\beta}_t.
\end{equation}

The effective drift $g(X_t,t)$ of Eq.~\eqref{appeq:sde_q} may be obtained from the solution of the variational problem of minimising the free energy
\begin{equation}\label{eq:free_energy}
    \mathcal{F}[Q] = \mathcal{KL}\Big(Q(X_{0:T})||\PP(X_{0:T}\mid \hat{f}) \Big)- \sum \limits_{k=1}^K {E}_Q[ \ln \PP(\mathcal{O}_{k}\mid X_{t_k})].
\end{equation}

By applying the Cameron-Girsanov-Martin theorem we can express the Kullback-Leibler divergence between the two path measures induced by the diffusions with drift $\hat{f}(x)$ and $g(x,t)$ as
\begin{align}
    \mathcal{KL}\Big(Q(X_{0:T})||\PP(X_{0:T}|\hat{f}) \Big) &= E_{{Q}}\left[\text{ln}\left(\frac{d {Q}(X_{0:T})}{d {P} \left( X_{0:T} \vert \hat{f}  \right)} \right)\right] \\ 
    &=E_{{Q}}\left[  \left( - \frac{1}{2}  \int_0^T { { \|\hat{f}(X_t)-g(X_t,t) \|_{\sigma^{2}}^2}\text{d}t} + \int_0^T {\frac{ \hat{f}(X_t)-g(X_t,t)  }{{\sigma^{2}}} \text{d}\boldsymbol{\beta}_t} \right)\right]\\
    &=E_{{Q}}\left[  \left( - \frac{1}{2}  \int_0^T { { \|\hat{f}(X_t)-g(X_t,t) \|_{\sigma^{2}}^2}\text{d}t} +V_T \right)\right]\\
    &= \frac{1}{2} \int \limits^T_0 \int \| g(x,t) -\hat{f}(x)   \|_{\sigma^{2}}^2 \, q_t(x)\, \text{d}x \, \text{d}t + \mathfrak{C} \label{eq:KL_1},
\end{align}
where $q_t(x)$ stands for the marginal density for $X_t$ of the approximate process. In the third line we have introduced the random variable $V_T = \int_0^T {\frac{ \hat{f}(X_t)-g(X_t,t)  }{{\sigma^{2}}} \text{d}\boldsymbol{\beta}_t}$. Under the assumption that the function ${\ell(X_t) = \hat{f}(X_t)-g(X_t,t)}$ is bounded, piece-wise continuous, and in $L^2[0,\infty)$ , $V_T$ follows the distribution $\mathcal{N}\left(V_T \mid 0, \int_0^T \ell^2(s) \text{d}s\right)$, which for a given $T$ will result into a constant $\mathfrak{C}$. Thus the second term in Eq.~\eqref{eq:KL_1} is not relevant for the minimisation of the free energy and will be omitted. 

We can thus express the free energy of Eq.~\eqref{eq:free_energy} as~\cite{opper2019variational}
\begin{equation} \label{appeq:free_energy2}
    \mathcal{F}[Q] = \frac{1}{2} \int \limits^T_0 \int \Big[ \|g(x,t) - \hat{f}(x)\|_{\sigma^{2}}^2 + U(x,t)   \Big] \, q_t(x)\, \text{d}x \,\text{d}t,
\end{equation}
where the term $U(x,t)$ accounts for the observations $U(x,t) = - \sum \limits_{t_k} \ln \PP(\mathcal{O}_k \mid x) \delta(t- t_k)$.

The minimisation of the functional of the free energy can be construed as a stochastic control problem~\cite{opper2019variational} with the objective to identify a time-dependent drift adjustment $u(x,t):=g(x,t) - \hat{f}(x)$ for the system with drift $\hat{f}(x)$ so that the controlled dynamics fulfil the constraints imposed by the observations.

For the case of exact observations, i.e., for an observation process $\psi(x) = x$, we can compute the drift adjustment for each of the $K-1$ inter-observation intervals independently. Thus for each interval between consecutive observations, we identify the optimal control $u(x,t)$ required to construct a stochastic bridge following the dynamics of Eq.~\eqref{appeq:prior_sde} with initial and terminal states the respective observations $\mathcal{O}_k$ and $\mathcal{O}_{k+1}$. 

The optimal drift adjustment for such a stochastic control problem for the inter-observation interval between $\mathcal{O}_k$ and $\mathcal{O}_{k+1}$ can be obtained from the solution of the backward equation (see~\cite{maoutsa2022deterministic, maoutsa2021deterministica})
\begin{equation}
    \frac{\partial \phi_t(x)}{\partial t} = - \mathcal{L}_{\hat{f}}^{\dagger} \phi_t(x) + U(x,t) \phi_t(x),
\end{equation}
with terminal condition $\phi_T(x) = \chi(x) = \delta(x-\mathcal{O}_{k+1}) $ and with $\mathcal{L}_{\hat{f}}^{\dagger}$ denoting the adjoint Fokker-Planck operator for the process of Eq.~\eqref{appeq:prior_sde}.
As shown in Maoutsa et al.~\cite{maoutsa2022deterministic, maoutsa2021deterministica} the optimal drift adjustment $u(x,t)$ can be expressed in terms of the difference of the logarithmic gradients of two probability flows
\begin{equation}
    u^*(x,t) = D \Big( \nabla \ln q_{T-t}(x) - \nabla \ln \rho_t(x) \Big),
\end{equation}
where $\rho_t$ fulfils the forward (filtering) partial differential equation (PDE)
 \begin{equation}
\frac{\partial \rho_t(x)}{\partial t} = {\cal{L}}_{\hat{f}} \rho_t(x) - U(x,t) \rho_t(x),
\label{eq:FPE2} 
\end{equation}
while $q_t$ is the solution of a time-reversed PDE that depends on the logarithmic gradient of $\rho_t(x)$
\begin{align}\label{Fokker_bridge3}
\frac{\partial {q}_{t}(x)}{\partial t} &= 
-\nabla\cdot \Bigg[\Big(\sigma^2\nabla \ln  \rho_{T-t} (x)  - f(x, T-t)\Big)  {q}_{t} (x)\Bigg] +  \frac{\sigma^2}{2} \nabla^2 {q}_{t} (x) , 
\end{align}
with initial condition ${q}_{0} (x) \propto \rho_T(x) \chi(x)$
.

\subsubsection{Approximate posterior over paths \underline{with} geometric constraints.}\label{appsec:with}

The previously described construction of the approximate measure in terms of stochastic bridges is relevant when the observations have non vanishing probability under the law of the prior diffusion process of Eq.~\eqref{appeq:prior_sde}. However, when the prior process (with the estimated drift $\hat{f}$) differs considerably from the process that generated the observations, such a construction might either provide a bad approximation of the underlying path measure, or show slow numerical convergence in the construction of the diffusion bridges.
To overcome this issue, we consider here additional constraints for the posterior process that force the paths of the posterior measure to respect the local geometry of the observations. In the following we provide a brief introduction on the basics of Riemannian geometry and consequently continue with the geometric considerations of the proposed method.

\paragraph{Riemannian geometry.}
A $d$-dimensional \textbf{Riemannian manifold}~\cite{do1992riemannian,lee2018introduction} $\left(\mathcal{M}, h \right)$ embedded in a $D$-dimensional ambient space $\mathcal{X} = \mathcal{R}^D$ is a smooth curved $d$-dimensional surface 
endowed with a smoothly varying inner product (Riemannian) \textbf{metric} $h: x \rightarrow \langle \cdot | \cdot \rangle_x$ on $\mathcal{T}_x\mathcal{M}$. A tangent space $\mathcal{T}_x \mathcal{M}$ is defined at each point $x \in \mathcal{M}$. The Riemannian metric $h$ defines a canonical volume measure on the manifold $\mathcal{M}$. Intuitively this characterises how to compute inner products locally between points on the tangent space of the manifold $\mathcal{M}$, and therefore determines also how to compute norms and thus distances between points on $\mathcal{M}$.

A \textbf{coordinate chart} $(G,\phi)$ provides the mapping from an open set $G$ on $\mathcal{M}$ to an open set $V$ in the Euclidean space. The dimensionality of the manifold is $d$ if for each point $x\in \mathcal{M}$ there exists a local neighborhood $G \subset  \mathcal{R}^d$.
We can represent the metric $h$ on the local chart $(G,\phi)$ by the positive definite matrix (\textbf{metric tensor}) $H(x) = (h_{i,j})_{x, 0 \leq i,j,\leq d} = \left( \langle  \frac{\partial}{\partial x_i}|  \frac{\partial}{\partial x_j}\rangle_x  \right)_{0 \leq i,j,\leq d}$ at each point $x \in G$.

For $v,w \in \mathcal{T}_x\mathcal{M}$ and $x \in G$, their inner product can be expressed in terms of the matrix representation of the metric  $h$ on the tangent space $\mathcal{T}_x\mathcal{M}$ as $\langle v|w \rangle_x = v^{\top} H(x)w$, where $H(x)\in \mathcal{R}^{d \times d}$ .

The \textbf{length of a curve} $\gamma:[0,1]\rightarrow \mathcal{M}$ on the manifold is defined as the integral of the norm of the tangent vector 
\begin{equation}\label{eq:ell}
\ell(\gamma_{t'}) = \int^1_0\| \dot{\gamma}_{t'}\|_h \text{d}t' = \int^1_0 \sqrt{ \dot{\gamma}_{t'}^{\top} H(\gamma_{t'}) \dot{\gamma}_{t'}    } \text{d}t',\end{equation}
where the dotted letter indicates the velocity of the curve $\dot{\gamma}_{t'}=\partial_{t'} \gamma_{t'}$. A \textbf{geodesic curve} is a locally length minimising smooth curve that connects two given points on the manifold.

\paragraph{Riemannian geometry of the observations.} 
For approximating the posterior over paths we take into account the geometry of the invariant density as it is represented by  the observations.  
To that end, we consider systems whose dynamics induce invariant (inertial) manifolds that contain the global attractor of the system and on which system trajectories concentrate~\cite{wiggins1994normally,mohammed1999stable, girya1995inertial, fenichel1971persistence, arnold1990stochastic, carverhill1985flows}. We assume thus that the continuous-time trajectories $X_{0:T} \in \mathcal{R}^d$ of the underlying system concentrates on an invariant manifold $\mathcal{M} \in \mathcal{R}^{m \leq d}$ of dimensionality $m$ (possibly) smaller than $d$.
The discrete-time observations $\mathcal{O}_k$ are thus samples of the manifold $\mathcal{M}$.
The central premise of our approach is that \textbf{unobserved paths between successive observations will be lying either \emph{on} or \emph{in the vicinity} of the manifold} $\mathcal{M}$. In particular, we postulate that unobserved paths should lie \textbf{in the vicinity of geodesics that connect consecutive observations} on $\mathcal{M}$. To that end we propose a path augmentation framework that constraints the augmented paths to lie in the vicinity of identified geodesics between consecutive observations.

However, while this view of a lower dimensional manifold embedded in a higher dimensional ambient space helps to build our intuition for the proposed method, for computational purposes we adopt a complementary view inspired by the discussion in~\cite{frohlich2021bayesian}. According to this view, we consider the entire observation space $\mathcal{R}^d$ as a smooth Riemannian manifold, $\mathcal{M}\dot{=}\mathcal{R}^d$, characterised by a Riemannian metric $h$. The effect of the nonlinear geometry of the observations is then captured by the metric $h$. Thus to approximate the geometric structure of the system's invariant density, we learn the Riemannian metric tensor $H:\mathcal{R}^d \rightarrow \mathcal{R}^{d \times d} $ and compute the geodesics between consecutive observations according to the learned metric. Intuitively according to this view the observations $\{\mathcal{O}_k\}^K_{k=1}$ introduce distortions in the way we compute distances on the state space.

In effect this approach does not reduce the dimensionality of the space we operate, but changes the way we compute inner products and thus distances, lengths, and geodesic curves on $\mathcal{M}$. The alternative perspective of working on a lower dimensional manifold would strongly depend on the correct assessment of the dimensionality of said manifold. For example, one could use a Variational Autoencoder to approximate the observation manifold and subsequently obtain the Riemannian metric from the embedding of the manifold mediated by the decoder. 
However, our preliminary results of such an approach revealed that such a method requires considerable fine tuning to adapt to the characteristics of each dynamical system and is sensitive to the estimation of the dimensionality of the approximated manifold. 

To learn the Riemannian metric and compute the geodesics we follow the framework proposed by Arvanitidis et al. in~\cite{arvanitidis2019fast}.
In particular, we approximate the local metric induced by the observations at location $\mathbf{x}$ of the state space, in a non-parametric form by the inverse of the weighted local diagonal covariance computed on the observations as~\cite{arvanitidis2019fast}
\begin{equation}
    H_{dd}(\mathbf{x}) = \left(  \sum\limits^K_{i=1} w_i(\mathbf{x}) \left( x^{(d)}_i - x^{(d)}\right)^2 + \epsilon   \right)^{-1},
\end{equation}
with weights $w_i(\mathbf{x}) = \exp \left(- \frac{\|  \mathbf{x}_i - \mathbf{x} \|^2_2}{2 \sigma^2_{\mathcal{M}}}  \right)$, and $x^{(d)}$ denoting the $d$-th dimensional component of the vector $\mathbf{x}$. The parameter $\epsilon > 0$ ensures non-zero diagonals of the weighted covariance matrix, while $\sigma_{\mathcal{M}}$ characterises the curvature of the manifold.

Between consecutive observations for each interval $[\mathcal{O}_k, \mathcal{O}_{k+1}]$, we identify the geodesic $\gamma^k_{t'}$ as the energy minimising curve, i.e., as the minimiser of the kinetic energy functional $\mathcal{E}(\gamma^k_{t'}) =\int^1_0 L_{{\mathcal{M}}}(\gamma^k_{t'}, \dot{\gamma}^k_{t'}) \text{d}t'$
\begin{equation} \nonumber
  \gamma^{k*}_{t'} =  \underset{\gamma^k_{t'}, \gamma^k_0 = \mathcal{O}_k, \gamma^k_1=\mathcal{O}_{k+1}}{\arg\min} \int^1_0 L_{{\mathcal{M}}}(\gamma^k_{t'}, \dot{\gamma}^k_{t'}) \text{d}t',
\end{equation}

\begin{equation} \label{eq:machlup}
\text{with} \;\;\;\;  \int^1_0 L_{{\mathcal{M}}}(\gamma^k_{t'}, \dot{\gamma}^k_{t'}) \text{d}t'= \frac{1}{2}  \int^1_0 \|\dot{\gamma}^k_{t'} \|^2_h  ,
\end{equation} \label{eq:geodesic_lagrangian}
where $L_{{\mathcal{M}}}(\gamma^k_{t'}, \dot{\gamma}^k_{t'})$ denotes the Lagrangian.
The minimising curve of this functional is the same as the minimiser of the curve length functional $\ell(\gamma_{t'})$ (Eq.~\eqref{eq:ell}), i.e., the geodesic~\cite{do1992riemannian}.

By applying calculus of variations, the minimising curve of the functional $\mathcal{E}(\gamma^k_{t'}) $ can be obtained from the Euler-Lagrange equations, resulting in the following system of second order differential equations~\cite{arvanitidis2017latent,do1992riemannian}
\begin{equation}\label{eq:geode}
    \ddot{\gamma_t}^k = -\frac{1}{2} {H(\gamma^k_t)}^{-1} \Bigg( 2 \left( I \otimes (\dot{\gamma_t}^k)^{\top} \right) \frac{\partial \text{vec}[H(\gamma^k_t)]}{\partial \gamma^k_t}  \dot{\gamma_t}^k  - \frac{\partial \text{vec}[H(\gamma^k_t)]^{\top}}{\partial \gamma^k_t} \left(\dot{\gamma_t}^k \otimes \dot{\gamma_t}^k  \right)\Bigg),
\end{equation}
with boundary conditions $\gamma^k_0 = \mathcal{O}_k $ and $ \gamma^k_1=\mathcal{O}_{k+1}$,
where $\otimes$ stands for the Kroenecker product, and $\text{vec}[A]$ denotes the vectorisation operation of matrix $A$ through stacking the columns of $A$ into a vector.
Arvanitidis et al.~\cite{arvanitidis2019fast} obtain the geodesics by approximating the solution of the boundary value problem of Eq.~\eqref{eq:geode} with a probabilistic differential equation solver.

\paragraph{Extended free energy functional.} We denote the collection of individual geodesics by $\Gamma_t\dot{=} \{\gamma^k_{t'}\}_{t=(k-1)\tau+t' \tau}$, where $\gamma^k_{t'}$ is the geodesic connecting $\mathcal{O}_k$ and $\mathcal{O}_{k+1}$, and $t'\in [0,1]$ denotes a rescaled time variable. Additional to the constraints imposed in the previously explained setting (Sec~\ref{appsec:without}), here we add an extra term in the free energy ${U_{\mathcal{G}}(x,t) \dot{=} \| \Gamma_t -x \|^2 }$ that accounts for the local geometry of the invariant density, and guides the latent path towards the geodesic curves $\gamma^k_{t'}$ that connect consecutive observations 
\begin{equation} \label{apeq:free_energy2}
 \small   \mathcal{F}[Q] = \frac{1}{2} \int \limits^T_0 \int \Big[ \|g(x,t) - \hat{f}(x)\|_{\sigma^2}^2 + U_{\mathcal{O}}(x,t) + \beta U_{\mathcal{G}}(x,t)   \Big] \, q_t(x)\, dx \,\text{d}t.
\end{equation}
Here we denote the observation term by $U_{\mathcal{O}}(x,t) \dot{=} - \sum_{t_k} \ln \PP(\mathcal{O}_k|x) \delta(t- t_k)$, while $\beta$ stands for a weighting constant that determines the relative weight of the geometric term in the control objective.

 Following~\cite{opper2019variational}, for each inter-observation interval $[\mathcal{O}_k, \mathcal{O}_{k+1}]$ we identify the posterior path measure (minimiser of Eq.~\eqref{apeq:free_energy2}) by the solution of a stochastic optimal control problem~\cite{maoutsa2022deterministic} with the objective to obtain a time-dependent drift adjustment $u(x,t):=g(x,t) - \hat{f}(x)$ for the system with drift $\hat{f}(x)$ with initial and terminal constraints defined by $U_{\mathcal{O}}(x,t)$, and additional path constraints $U_{\mathcal{G}}(x,t)$.


\if False
\paragraph{Most probable path between observations.}
We identify the most probable path between observations
in the Onsager-Machlup sense~\cite{fujita1982onsager} by the curve $\gamma^*_t:[0,1] \rightarrow \widehat{\mathcal{M}}$ with $\gamma_0=O_k$ and $\gamma_1=O_{k+1}$ lying in the vicinity of the geodesic curve between the two endpoints. More precisely the most probable path $ \gamma^*_t$ minimises the functional $\int^1_0 L_{\widehat{\mathcal{M}}}(\gamma_t, \dot{\gamma}_t) \text{d}t$~\cite{kuhnel2019differential, sommer2015anisotropic}

\noindent\begin{minipage}{.4\linewidth}
\begin{equation} \nonumber
  \gamma^*_t =  \underset{\gamma_t, \gamma_0 = \mathcal{O}_k, \gamma_1=\mathcal{O}_{k+1}}{\arg\min} \int^1_0 L_{\widehat{\mathcal{M}}}(\gamma_t, \dot{\gamma}_t) \text{d}t, 
\end{equation}
\end{minipage}%
\begin{minipage}{.6\linewidth}
\begin{equation} \label{eq:machlup}
\text{with} \;\;\;\;  \int^1_0 L_{\widehat{\mathcal{M}}}(\gamma_t, \dot{\gamma}_t) \text{d}t= - \frac{1}{2} \int^1_0 \|\dot{\gamma}_t \|^2_h + \frac{1}{6} \mathcal{S}(\gamma_t)  \text{d}t ,
\end{equation}
\end{minipage}

where $\mathcal{S}(\gamma_t)$ stands for the scalar curvature of the manifold $\widehat{\mathcal{M}}$, while the first term captures the geodesic energy. 

\fi

\subsection{Approximate posterior over drift functions.}\label{appsec:drift_inference}

For a fixed path measure ${Q}$, the optimal measure for the drift ${Q}_f$ is a Gaussian process given by
\begin{equation} \label{appeq:drift_measure}
    {Q}_f \propto \PP_f \exp\left({ -\frac{1}{2} \int  \|f(x)\|_{\sigma^{2}}^2 A(x) - 2 \langle f(x), B(x) \rangle_{\sigma^{2}}  } \text{d}x\right),
\end{equation}
with $$A(x)\dot{=} \int^T_{0} p_t(x) \text{d}t,$$ and $$B(x)\dot{=} \int^T_{0} p_t(x) g(x,t) \text{d}t, $$ where $p_t(x)$ denotes the marginal constrained density of the state $X_t$. The function $g(x,t)$ denotes the effective drift.

 We assume a Gaussian process prior for the unknown function $f$, i.e., $f \sim \PP_0({f}) =\text{GP}(m^f, k^f)$ where $m^f$ and $k^f$ denote the mean and covariance function of the Gaussian process. Following Ruttor \emph{et al.}~\cite{ruttor2013approximate}, we employ a sparse kernel approximation for the drift $f$ by optimising the function values over a sparse set of $S$ inducing points $\{Z_i\}^{S}_{i=1}$.
We obtain the resulting drift from
\begin{equation}
    \hat{f}_S(x) = k^f(x,\mathcal{Z}) \left( I + \Lambda \, \mathcal{K}_S  \right)^{-1} \mathbf{d},
\end{equation}
where we have defined introduced the notation $\mathcal{K}_S \dot{=} k^f(\mathcal{Z},\mathcal{Z}) $
\begin{equation}
    \Lambda = \frac{1}{\sigma^2} \mathcal{K}^{-1}_S \left(   \int k^f(\mathcal{Z},x) A(x) k^f(x,\mathcal{Z}) \text{d}x \right)   \mathcal{K}^{-1}_S.
\end{equation}

\begin{equation}
    \mathbf{d} = \frac{1}{\sigma^2} \mathcal{K}^{-1}_S \left(   \int k^f(\mathcal{Z},x) B(x) \text{d}x \right)   \mathcal{K}^{-1}_S,
\end{equation}

The associated variance results similarly from the equation
\begin{equation}
    \Sigma^2_S (x) = k^f(x,x) - k^f(x,\mathcal{Z}) \left( I + \Lambda \, \mathcal{K}_S  \right)^{-1} \Lambda\, k^f(\mathcal{Z},x).
\end{equation}

We employ a sample based approximation of the densities in Eq.~\eqref{appeq:drift_measure} resulting from the particle sampling of the path measure $Q$.
Thus by representing the densities by samples, we can rewrite the density $p_{t}(x)$ in terms of a sum of Dirac delta functions centered around the particles positions
$$p_{t}(x) \approx \frac{1}{N} \sum^N_{j=1} \delta(x - X_j(t)),$$ and replace the Riemannian integrals with summation over particles. Here $X_j(t)$ represents the position of the $j$-th particle at time point $t$.

\section{Theoretical evidence that may support the use of geodesics as geometric constraints} \label{app:OM}

The Onsager-Machlup functional for diffusion processes has been known in theoretical physics as a characteriser of the most probable path (MPP) between two pre-defined states of the process. In~\cite{onsager1953fluctuations}, Onsager and Machlup used the thermal fluctuations of a diffusion process to show that the probability density of a path $\gamma \in C^1 \left([0,T], \mathcal{R}^d \right)$ in $\mathcal{R}^d$ over finite interval can be expressed as a Boltzmann factor 
\begin{equation}
\PP(\gamma) \sim \exp \left[ - \int^{T}_{0} L(\gamma(t), \dot{\gamma}(t) ) dt \right],
\end{equation}
where \begin{equation}L(\gamma(t), \dot{\gamma}(t) ) = \frac{1}{2} \| \frac{\dot{\gamma}(t) - f(\gamma(t))}{\sigma}\|^2 + \frac{1}{2} \nabla \cdot f(\gamma(t)).\footnote{Onsager and Machlup's initial work concentrated around linear processes and therefore the functional initially introduced by the did not include the second term with the divergence of $f$ as this is a constant for linear $f$. It was later added to the OM function to account for trajectory entropy corrections~\cite{taniguchi2007onsager,adib2008stochastic}}\end{equation}
The function $L(\gamma(t), \dot{\gamma}(t) )$ is known as the \textbf{Onsager-Machlup} function (action), while its integral over time is known as Onsager-Machlup action functional. It has been used as Lagrangian in Euler-Lagrange minimisation schemes to identify the \textbf{most probable path (MPP)} of a diffusion process between two given points in the state space~\cite{graham1977path, stratonovich1971probability}.

Stratonovich~\cite{stratonovich1971probability} considered the probability that a sample of a multidimensional diffusion process will lie in the vicinity of (within a tube of infinitesimal thickness around) an idealised smooth path in the state space. To compute this probability he constructed a probability functional which is identical to the Onsager-Machlup functional considered as Lagrangian for the diffusion process.
 Duerr et al.~\cite{durr1978onsager} considered scalar diffusion processes and constructed the Onsager-Machlup function from the asymptotic limit of the transition probability between the starting and end state of the path using a Girsanov transformation. 

\if False
Fujita et al.~\cite{fujita1982onsager} have shown that the log probability of a tube of width $\epsilon \rightarrow 0$ in the state space has limiting value value 
\begin{equation}
    \log \PP(\gamma) \rightarrow \mathfrak{C}_1 + \mathfrak{C}_2/\epsilon^2 - \int^T_0 \frac{1}{2} \| \dot{\gamma}(t) \|^2 dt,
\end{equation}
where the term $\| \dot{\gamma}(t) \|^2$ can be identified as the energy of the path, and $\mathfrak{C}_1$,  $\mathfrak{C}_2$ denote constants. This already bares some similarities with our approach, since the curve that minimises the energy functional along a path in path space can be identified as a geodesic curve.
\fi

Considering Brownian motions defined on  a Riemannian manifold $(\mathcal{M}, g)$ with associated Riemannian metric $g$, the Onsager-Machlup functional can be expressed as the integral over the Lagrangian~\cite{takahashi1981probability,graham1980onsager,grong2022most}
\begin{equation} \label{eq:OM_manifold}
    L(\gamma, \dot{\gamma}) = \frac{1}{2} \| \dot{\gamma}(t) \|_g^2 - \frac{1}{12} S(\gamma(t)),
\end{equation}
where $\| \cdot \|_g$ denotes the Riemannian norm on the tangent space $\mathcal{T}_X \mathcal{M}$ of the manifold with respect to the metric $g$, and $S(\cdot)$ stands for the scalar curvature of the manifold at each point. The first term is the Lagrangian used to identify geodesic curves on manifolds (c.f. Eq.~\eqref{eq:geodesic_lagrangian})

In our proposed formalism, for computational purposes we have assumed the entire $\mathcal{R}^d$ as smooth manifold. We can identify the first term of Eq.~\eqref{eq:OM_manifold} with the Lagrangian we optimised for computing the geodesics on the manifold $(\mathcal{R}^d, g)$, where $g$ is the metric learned from the observations.

However the system we observed was a diffusion process defined in $\mathcal{R}^d$ with an Euclidean metric. Constructing a path augmentation scheme that guides the augmented paths towards the geodesics of a diffusion defined with respect to a different metric raises questions about the validity of our approach. Here we should note that diffusions with a general state dependent diffusion coefficient $\sigma \in \mathcal{R}^{d \times m}$, and $m$-dimensional Brownian motion, can be considered as evolving on the manifold $\left(\mathcal{R}^{d}, g \right)$, with the associated metric $g = \left(\sigma \sigma^{\top} \right)^{-1}$~\cite{capitaine2000onsager}. Thus it may be possible to associate the metric learned from the data with the metric arising from a state dependent diffusion by applying a transformation akin to an inverse Lamperti transform~\cite{oksendal2003stochastic} to transform our learned SDE to one that would have induced the learned metric due to the state dependent diffusion. The existence of such a transformation would justify the proposed method. Our empirical results demonstrate that such a transformation may be possible.



\section{Does the proposed approach invalidate the Markovian property of the diffusion process?}
The proposed path augmentation seemingly invalidates the Markovian property of the diffusion process.
According to the Markov property of the diffusion of Eq.~\eqref{eq:system}, the system state $X_{k \tau+\delta t}$ should depend only the state $X_{k \tau}$, i.e., the observation $\mathcal{O}_k$. The proposed augmentation makes the state $X_{k \tau+\delta t}$ depending not only on the next observation $\mathcal{O}_{k+1}=X_{(k+1) \tau}$, but also on past and future states that lie in the vicinity of these observations. 

We effectively construct the augmented paths to compute the likelihood of a drift estimate. To compute this likelihood we require to evaluate the transition probabilities between consecutive observations. Since for general nonlinear systems the transition probabilities are in general intractable, we have to resort to numerical approximations. Ideally we would approximate the transition density with a bridge sampler that would consider the nonlinear estimated SDE conditioned to pass though consecutive observations. However for coarse drift estimates, the observations have zero probability under the law of the estimated SDE, and construction of those bridges would result either in very taxing computations or would fail altogether. Instead, here, we compute the likelihood of a "corrected" estimate (the correction resulting from the invariant density) under which the observations have non-zero probability, and subsequently re-estimate the drift on the augmented path with this "corrected" estimate.
 By taking into account the local geometry of the observations, we provide systematic corrections for the misestimated drift function to generate the augmented paths. This effectively nudges the augmentation process towards the second observation of each inter-observation interval through the path constraint that forces the augmented paths towards the geodesics.

 \section{Details on numerical experiments}\label{sec:details}
 We simulated a two dimensional Van der Pol oscillator
 with drift function 
 \begin{align}
        f_1(x,y) &= \mu(x - \frac{1}{3}x^3-y)\\
        f_2(x,y) &= \frac{1}{\mu}x,
 \end{align}
 starting from initial condition $ x0 = [1.81, -1.41]$
and under noise amplitudes $\sigma=\{0.25, 0.50, 0.75, 1.00\}$ for total duration of $T=\{500, 1000\}$ time units. The employed inter-observation intervals $\tau=\{80, 120, 160, 200, 240, 280,  320\}$. The last inter-observation interval exceeds the half period of the oscillator and thus samples only a single state per period. This resulted in erroneous estimates. 
In this setting this indicates the upper limit of $\tau$ for which we can provide estimates. However for any inference method, if the observation process samples only one observation per period, identifying the underlying force field without additional assumptions is not possible with temporal methods. The discretisation time-step used for simulation of the ground truth dynamics, and path augmentation $\delta t=0.01$.
For sampling the controlled bridges we employed $N=100$ particles evolving the associated ordinary differential equation as described in~\cite{maoutsa2022deterministic,maoutsa2020interacting}. The logarithmic gradient estimator used $M=40$ inducing points. The sparse Gaussian process for estimating the drift was based on a sparse kernel approximation of $S=300$ points. In the presented simulation we have employed a weighting parameter $\beta = 0.5$ (Eq.~\eqref{apeq:free_energy2}). This provides a moderate pull towards the invariant density. The example in Figure~\ref{fig:path_augmentations} was constructed with $\beta=1$ and provides a better approximation of the transition density, than $\beta=0.5$.
\end{document}
